\documentstyle[12pt,psfig]{article}
\newcommand{\nsect}{\setcounter{equation}{0}
\def\theequation{\thesection.\arabic{equation}}\section}

\catcode`\@=11
\def\marginnote#1{}
\def\ifmath#1{\relax\ifmmode #1\else $#1$\fi}

\def\msTop{m_{\,\widetilde{T}}}
\def\mstop{m_{\,\widetilde{t}}}
\def\mstopeff{m_{\,\widetilde{t}}^{\rm eff}}

\def\bold#1{\setbox0=\hbox{$#1$}%
     \kern-.025em\copy0\kern-\wd0
     \kern.05em\copy0\kern-\wd0
     \kern-.025em\raise.0433em\box0 }

\def\GENITEM#1;#2{\par\vskip6pt \hangafter=0 \hangindent=#1
   \Textindent{$ #2$ }\ignorespaces}

\newcount\hour
\newcount\minute
\newtoks\amorpm
\hour=\time\divide\hour by60
\minute=\time{\multiply\hour by60 \global\advance\minute by-
\hour}
\edef\standardtime{{\ifnum\hour<12 \global\amorpm={am}%
    \else\global\amorpm={pm}\advance\hour by-12 \fi
    \ifnum\hour=0 \hour=12 \fi
    \number\hour:\ifnum\minute<100\fi\number\minute\the\amorpm}}
\edef\militarytime{\number\hour:\ifnum\minute<100\fi\number\minute}
\def\draftlabel#1{{\@bsphack\if@filesw {\let\thepage\relax
  \xdef\@gtempa{\write\@auxout{\string
    \newlabel{#1}{{\@currentlabel}{\thepage}}}}}\@gtempa
    \if@nobreak \ifvmode\nobreak\fi\fi\fi\@esphack}
     \gdef\@eqnlabel{#1}}
\def\@eqnlabel{}
\def\@vacuum{}
\def\draftmarginnote#1{\marginpar{\raggedright\scriptsize\tt#1}}
\def\draft{\oddsidemargin -.5truein
        \def\@oddfoot{\sl preliminary draft \hfil
        \rm\thepage\hfil\sl\today\quad\militarytime}
        \let\@evenfoot\@oddfoot \overfullrule 3pt
        \let\label=\draftlabel
        \let\marginnote=\draftmarginnote

\def\@eqnnum{(\theequation)\rlap{\kern\marginparsep\tt\@eqnlabel}%
\global\let\@eqnlabel\@vacuum}  }
\def\preprint{\twocolumn\sloppy\flushbottom\parindent 1em
        \leftmargini 2em\leftmarginv .5em\leftmarginvi .5em
        \oddsidemargin -.5in    \evensidemargin -.5in
        \columnsep 15mm \footheight 0pt
        \textwidth 250mmin      \topmargin  -.4in
        \headheight 12pt \topskip .4in
        \textheight 175mm
        \footskip 0pt

\def\@oddhead{\thepage\hfil\addtocounter{page}{1}\thepage}
        \let\@evenhead\@oddhead \def\@oddfoot{} \def\@evenfoot{}
}
\def\titlepage{\@restonecolfalse\if@twocolumn\@restonecoltrue\onecolumn
     \else \newpage \fi \thispagestyle{empty}\c@page\z@
        \def\thefootnote{\fnsymbol{footnote}} }
\def\endtitlepage{\if@restonecol\twocolumn \else  \fi
        \def\thefootnote{\arabic{footnote}}
        \setcounter{footnote}{0}}  %\c@footnote\z@ }
\catcode`@=12
\relax
\def\be{\begin{equation}}
\def\ee{\end{equation}}
\def\bea{\begin{eqnarray}}
\def\eea{\end{eqnarray}}
\def\simlt{\stackrel{<}{{}_\sim}}
\def\simgt{\stackrel{>}{{}_\sim}}

\def\NPB#1#2#3{{\it Nucl.~Phys.} {\bf{B#1}} (19#2) #3}
\def\PLB#1#2#3{{\it Phys.~Lett.} {\bf{B#1}} (19#2) #3}
\def\PRD#1#2#3{{\it Phys.~Rev.} {\bf{D#1}} (19#2) #3}
\def\PRL#1#2#3{{\it Phys.~Rev.~Lett.} {\bf{#1}} (19#2) #3}

\def\MPLA#1#2#3{{\it Mod.~Phys.~Lett.} {\bf#1} (19#2) #3}

\def\AP#1#2#3{{\it Ann.~Phys.} {\bf#1} (19#2) #3}

\def\HPA#1#2#3{{\it Helv.~Phys.~Acta} {\bf#1} (19#2) #3}
\def\JETPL#1#2#3{{\it JETP~Lett.} {\bf#1} (19#2) #3}

\def\mst11{m_{\;\widetilde{t}_{1}}}

\def\mst22{m_{\;\widetilde{t}_{2}}}
\def\mst12{m_{\;\widetilde{t}_{1,2}}}

\def\msb11{m_{\;\widetilde{b}_{1}}}
\def\msb22{m_{\;\widetilde{b}_{2}}}
\def\msb12{m_{\;\widetilde{b}_{1,2}}}

\def\mtilde2{\widetilde{m}^{2}}

\relax
\begin{document}
\topmargin-2.5cm
%\draft
%\preprint
%
\begin{titlepage}
\begin{flushright}
CERN-TH/96-30\\
IEM-FT-126/96 \\
hep--ph/9603420 \\
\end{flushright}
\vskip 0.3in
\begin{center}{\Large\bf Opening the Window
for Electroweak  Baryogenesis}
\footnote{Work supported in part by the European Union
(contract CHRX/CT92-0004) and CICYT of Spain
(contract AEN95-0195).}
\vskip .5in
{\bf M. Carena~$^{\dagger}$},
{\bf M. Quir\'os~$^{\ddagger}$} and {\bf C.E.M. Wagner~$^{\dagger}$}
\vskip.35in
$^{\dagger}$~CERN, TH Division, CH--1211 Geneva 23, Switzerland\\
$^{\ddagger}$~Instituto de Estructura de la Materia, CSIC, Serrano
123, 28006 Madrid, Spain
\end{center}
\vskip1.3cm
\begin{center}
{\bf Abstract}
\end{center}
\begin{quote}
We perform an analysis of the behaviour of the
electroweak phase transition in the Minimal Supersymmetric Standard
Model, in the presence of light stops.
We show that, in previously unexplored regions of parameter space,
the order parameter $v(T_c)/T_c$ can become significantly 
larger than one, for values of the Higgs and supersymmetric particle masses
consistent with the present experimental bounds. This implies
that baryon number can be efficiently generated at the
electroweak phase transition.
As a by-product of this study, we present an analysis of
the problem of colour breaking minima at zero and finite temperature
and we use it to investigate
the region of parameter space preferred by the best fit
to the present precision electroweak measurement data, in
which the left-handed stops are much heavier than the right-handed ones.
\end{quote}
\vskip1.cm

\begin{flushleft}
CERN-TH/96-30\\
March 1996 \\
\end{flushleft}

\end{titlepage}
\setcounter{footnote}{0}
\setcounter{page}{0}
\newpage
%
% BODY
\nsect{Introduction}
The origin of the baryon asymmetry of the Universe remains  one
of the most intriguing open questions in
high energy physics~\cite{baryogenesis}. It was long assumed
that this question could only be answered by
the knowledge of the physics at
very short distances, of the order of the Grand
Unification or the Planck scale.
This general assumption was challenged through the
discovery that anomalous processes~\cite{anomaly}
can partially or totally
erase the baryon asymmetry generated at
extremely high energies~\cite{sphalerons}.
Much attention was hence devoted to
the possibility of generating the baryon asymmetry at the electroweak phase
transition~\cite{reviews},
assuming that no new physics beyond the Standard Model is present
at the weak scale.

The Standard Model has all the required properties for the
generation of the baryon
asymmetry:  CP violation, baryon number violating processes and,
in addition, non-equilibrium processes that are
generated at the first-order electroweak
phase transition.  To generate the required baryon asymmetry, the
electroweak phase transition
must be strongly first order. Quantitatively,
the requirement is that the ratio of the vacuum expectation
value of the Higgs field at the critical temperature to the critical
temperature must be larger than 1~\cite{first},
\be
\frac{v(T_c)}{T_c}  \simgt 1.
\label{orderp}
\ee
The Higgs potential of the Standard Model
at finite temperature can be given by
\be
V_{eff}^{\rm SM} = -m^2(T) \phi^2 - E_{\rm SM}
\;T\;\phi^3 + \frac{\lambda(T)}{2} \phi^4+\cdots,
\label{veff}
\ee
where the coefficient of the cubic term is
\be
\label{ESM}
E_{\rm SM}\sim \frac{2}{3}\
\left(\frac{2M_W^3+M_Z^3}{\sqrt{2} \pi v^3}\right)\; ,
\ee
$\langle\phi(T)\rangle = v(T)/\sqrt{2}$, and the
normalization of $\phi(T)$ is chosen  such
that its zero temperature vacuum
expectation value is $\langle\phi(T=0)\rangle = v/\sqrt{2}$,
with $v = 246.22$ GeV.
The critical temperature is defined as that one for which the
symmetry-breaking minimum has the same depth as
the symmetry-preserving one.
From Eq.~(\ref{veff}), it is easy to show that
\begin{equation}
\frac{v(T_c)}{T_c} \simeq\frac{\sqrt{2}\; E_{\rm SM}}{\lambda}.
\label{vot}
\end{equation}
The effective quartic coupling $\lambda$
at $T_c$ is closely related to its zero temperature
value, implying that the requirement
of Eq.~(\ref{orderp}) puts an upper
bound on the Higgs mass. This upper bound was estimated by the analysis
of the improved one-loop effective potential to
be of order 40 GeV~\cite{improvement}. It was
subsequently shown that higher-loop effects can enhance the strength
of the first-order phase transition~\cite{twoloop}.
The most recent non-perturbative studies~\cite{nonpert}
indicate that the real upper bound is still below the present
experimental bound on the Higgs mass, $m_H \geq 65$ GeV.

Furthermore, in the Standard Model the source of CP-violation is
associated with the CP-violating phase in the Cabbibo-Kobayashi-Maskawa
matrix. Any CP-violating process is suppressed by powers of $m_f/M_V$,
where $m_f$ are the light-quark masses and $M_V$ is the mass of the
vector bosons. It was shown that these suppression factors are
sufficiently strong
to severely restrict the possible baryon number generation~\cite{CPSM}.

Thus, to generate the observed baryon asymmetry of the Universe at the
electroweak phase transition, the presence of new physics at the weak
scale is required. An interesting possibility is 
that the new physics be given by the minimal supersymmetric extension
of the Standard Model (MSSM). In the MSSM, new sources of CP-violation
are present~\cite{CPMSSM,SPCP},
which can serve to avoid the strong Standard Model suppression
discussed above~\cite{CPMSSM}.
Preliminary results on the behaviour of the electroweak
phase transition within this model~\cite{early,mariano1,mariano2}
showed that the situation can only be
improved slightly in comparison with the Standard Model case. This improvement
was associated with the presence of light supersymmetric partners of the top
quark (stops) and small values of $\tan\beta$.

In this article, we shall show that, in previously unexplored
regions of parameter space, the phase transition can be  more strongly
first order than previously derived, without being in conflict with any
phenomenological constraint.  We shall  follow the formalism and
conventions of Refs.~\cite{mariano1,mariano2}, where
some technical details relevant for this presentation can be found.

\nsect{Light Top Squark Effects}

In the following, we shall 
explain the reason why, as it was already  observed
in Ref.~\cite{mariano1}, the presence of light stops can help in
enhancing the strength of the first-order phase transition.
We shall work in the limit
$m_A \gg M_Z$, which implies that only one Higgs doublet $\phi$ survives
at scales of
order $T_c$~\footnote{This case is favoured by the strength of the
phase transition~\cite{mariano1} and by precision electroweak
measurements in the low $\tan\beta$ regime~\cite{precision}.}.
We shall also
concentrate on the case that the light stop is predominantly right-handed,
implying that  $m_Q^2 \gg m_U^2,\ m_t^2$,
where $m_Q^2$ and $m_U^2$ are
the soft supersymmetry-breaking squared mass parameters of
the left- and right-handed stops, respectively,
and $m_t$ is the running top quark mass.  This hierarchy of masses
is naturally expected in the small $\tan\beta$ regime, if supersymmetry
is broken in the hidden sector~\cite{COPW,CW}.
Moreover, this range of parameters is
selected by the best fit to the precision electroweak
data within the MSSM~\cite{precision}.
Indeed, large values of $m_Q^2$ assure a small supersymmetric contribution
to the oblique corrections, while low or negative values of $m_U^2$
can help in enhancing
the value of $R_b$, particularly in the presence of light charginos,
with a dominant Higgsino component and close to
the present experimental bounds.

Within the above framework,
the stop masses are approximately given by
\begin{eqnarray}
\mstop^2 & \simeq &m_U^2 + D_R^2 + m_t^2(\phi) \left( 1  -
\frac{\widetilde{A}_t^2}{m_Q^2}
\right)
\nonumber\\
\msTop^2 & \simeq& m_Q^2 + D_L^2 + m_t^2(\phi) \left( 1 +
\frac{\widetilde{A}_t^2}{m_Q^2}
\right) ,
\end{eqnarray}
where $m_t(\phi)=h_t\sin\beta\phi$ is the top quark mass,
$\mstop$ and $\msTop$ are the lightest and heaviest stop masses,
$D_{R,L}^2$ are the $D$-term contributions 
to the right- and left-handed stop
squared masses, respectively, $h_t$ is the top quark Yukawa coupling
and $\widetilde{A}_t = A_t - \mu/\tan\beta$ is the effective
stop mixing mass parameter.
The heaviest stop leads to a relevant contribution to the zero-temperature
effective
potential, which can be absorbed in a redefinition of the
parameters $m^2$ and
$\lambda$ in Eq.~(\ref{veff}). The contribution of the heavy stop to
the quartic coupling is quite
significant, growing with the fourth power of the top quark
mass and logarithmically
with $m_Q$~\cite{Higgs,BERZ}. Large values of $m_Q$ have
hence the effect of
increasing the Higgs mass.
Although larger values of the Higgs mass are welcome to
avoid the experimental
bound, they necessarily lead to a weakening of the first-order 
phase transition. Indeed, the running Higgs mass is
given by
\begin{equation}
m_H^2 = \lambda v^2,
\label{mh2}
\end{equation}
and hence, any increase in the Higgs mass is associated with an
increase of the quartic coupling $\lambda$, yielding lower values
of $v(T)/T$. Therefore, very large values of $m_Q$, above a few TeV,
are disfavoured from  this point of view.

In the above discussion we have ignored the effect of
operators of dimension higher than 4 in the effective potential.
In the numerical computations, we include the full one-loop
effective potential \cite{BERZ}, which goes beyond the approximation of
Eq.~(\ref{veff})~\cite{CEQW}.  For consistency,
in the numerical evaluations, we neglect the two-loop
effects on the Higgs mass. In this case, the Higgs mass expressions
obtained in Ref.~\cite{CEQW} reduce to the ones presented in
Ref.~\cite{BERZ}, with the only difference that one-loop
D-terms have been included in our computation. Observe, however,
that the most important zero temperature two-loop contributions
can be absorbed in a redefinition of the quartic coupling $\lambda$
and hence, due to Eqs.~(\ref{vot}) and (\ref{mh2}),
they will not modify the upper bound on the Higgs mass. The genuine
two-loop finite-temperature contributions, instead, have a more
relevant effect, making  the phase transition
more strongly first order  (see Ref.~\cite{JoseR}). This effect goes
beyond the standard model contributions~\cite{twoloop} discussed above.

The finite-temperature effects of the heaviest stop are
exponentially suppressed and hence we shall ignore them in the
discussion below. (They are, however, kept in the numerical
evaluations.) The lightest stop, instead, plays an important role and we
shall single out its most relevant effects. We have used a 
finite-temperature expansion for it and checked that the latter does not
break down in our region of parameters.
The improved one-loop finite temperature effective potential is given by
\be
\label{potMSSM}
V_{\rm eff}^{\rm MSSM} = -m^2(T) \phi^2 - T \;
\left[E_{\rm SM}\; \phi^3 +  (2 N_c) \frac{\left(\mstop^{2} +
\Pi_R(T)\right)^{3/2}}
{12 \pi} \right] + \frac{\lambda(T)}{2} \phi^4+\cdots
\ee
where
\be
\label{polarization}
\Pi_R(T)=\frac{4}{9}g_3^2 T^2+
\frac{1}{6}h_t^2 \left[1+\sin^2\beta \left(1 - \widetilde{A}_t^2/m_Q^2
\right)\right]
T^2+\left(\frac{1}{3}-\frac{1}{18}|\cos 2\beta|\right)g'^2 T^2
\ee
is the finite temperature self-energy contribution to
the right-handed squarks (see section 4), $g_3$ is the strong
gauge coupling and $N_c = 3$
is the number of colours. We have included in
Eq.~(\ref{polarization}) the contribution of the Standard Model
fields and the light supersymmetric partners, in particular
charginos, neutralinos and light stops~\footnote{We shall work
in the case of sufficiently heavy  gluinos,
right-handed sbottoms and first and second
generation squarks, so that their contributions
to $\Pi_R$ are Boltzmann-suppressed.}.
Observe that, in general, as happens with the longitudinal
components of the gauge bosons, the lightest stop contribution to the
effective potential does not induce a cubic term. This is mainly related to
the fact that the \lq\lq effective finite-temperature  stop mass"
is not vanishing in the symmetric phase.
At $\phi = 0$, this effective mass is given by
\be
\left(\mstopeff \right)^2(\phi=0) = m_U^2 + \Pi_R(T).
\label{effmass}
\ee
The second term in Eq.~(\ref{effmass}) is positive, and hence a
small effective mass can only be obtained through a negative value
of the soft supersymmetry-breaking parameter $m_U^2$.

Negative values of $m_U^2$ can hence
enhance the strength of the phase transition, particularly if they
are close to those for which $\mstopeff =0$.
Indeed, if $\mstopeff (\phi=0,T=T_c) \simeq 0$,
the strength of the cubic term in the effective
potential receives a
contribution proportional to the cube of the top quark Yukawa coupling,
\be
E \simeq E_{\rm SM}+ \frac{h_t^3 \sin^3\beta \left(1 -
\widetilde{A}_t^2/m_Q^2\right)^{3/2}}{2 \pi},
\label{totalE}
\ee
where the first term is the Standard Model contribution
[$E_{\rm SM} \sim 0.018$]. Observe that, if $\mstopeff \simeq 0$,
the effective cubic term at $T_c$ can be nine times as large as the
Standard Model one. Since
$v(T_c)/T_c \propto E/\lambda$, an enhancement by a factor 9 of
$E$ implies that the
allowed Higgs mass values can be enhanced by as much
as a factor three. Hence, in
the case of zero mixing in the stop sector,
$\widetilde{A}_t \simeq 0$, and in the absence
of additional phenomenological constraints, the bound
on the Higgs mass within the
MSSM can be of the order of $100$ GeV.  Large values of $\widetilde{A}_t$,
instead, reduce the
induced cubic term coefficient and, 
for $\widetilde{A}_t \simeq \pm m_Q$,
the strength of the
first-order phase transition is of the order of the Standard Model one.

For values of the Higgs mass 
$m_H \simlt M_Z$ we should be concerned
by the validity of the perturbative expansion for the thermal
field theory. The usual argument in the Standard Model~\cite{twoloop},
which yields the condition $m_H^2\ll M_W^2$, goes as follows. In the
Standard Model the strength of the first-order phase transition
is mainly dominated by gauge bosons [see  Eq.~(\ref{ESM})].
Therefore, in the region near the symmetry-breaking minimum of
the potential (\ref{veff}), the value of the field is
$\phi\sim g^3 T/\lambda$. Each additional loop of gauge bosons
costs a factor of $g^2 T$ and the loop expansion parameter is
obtained dividing $g^2 T$ by the leading mass of the
problem, i.e. $M_W$. This gives the loop expansion parameter for the
thermal perturbation theory of the Standard Model, as
\be
\label{betaSM}
\beta_{\rm SM}\sim \frac{g^2
T}{M_W}\sim\frac{\lambda}{g^2}\sim\frac{m_H^2}{M_W^2} .
\ee
The condition $\beta_{\rm SM}\ll 1$ provides the aforementioned
condition on the Higgs mass.

In the MSSM, the strength of the phase transition at one-loop is
dominated by the light stops [as can be seen from Eq.~(\ref{totalE})].
Hence, as far as the phase transition is concerned, we can safely neglect
the electroweak gauge couplings and concentrate on the top
Yukawa and strong gauge couplings.
Using now Eqs.~(\ref{potMSSM}), (\ref{effmass})
and (\ref{totalE}) we can see that in the region near the symmetry
breaking minimum, $\phi\sim(h_t\sin\beta)^3 T/\lambda$.
Additional 
bosonic loops (on the one-loop light stop diagram) 
are dominated by the exchange of light stops and
Higgs bosons, with an energetic cost
$\sim (h_t\sin\beta)^2 T$ and by the exchange of gluons with
a cost~\footnote{We are considering here the case
which strengthens as much as possible the phase transition and,
therefore, leads to the largest possible values of the Higgs
mass: negligible mixing in the stop sector, i.e.
$\widetilde{A}_t /m_Q \simeq 0$, and heavy
gluinos~\cite{mariano1,mariano2}.} $\sim g_3^2 T^2$.
Since for the experimental range of the top quark mass
$h_t \sin\beta \simeq g_3$ and considering the relevant mass
of the problem to be  $\mstopeff\sim m_t$,
we can write the loop expansion parameter for the thermal theory
in the MSSM as
\be
\label{betaMSSM}
\beta_{\rm MSSM}\sim\frac{(h_t\sin\beta)^2\;T}{\mstopeff}\sim
\frac{\lambda}{(h_t\sin\beta)^2}\sim\frac{m_H^2}{m_t^2}.
\ee
In this way, the condition for the validity of the perturbative
expansion $\beta_{\rm MSSM}\ll 1$ leads to the bound on the
Higgs mass $m_H^2\ll m_t^2$. In the above we
have used Eq.~(\ref{totalE}), which is only valid if
$\mstopeff(\phi=0)$ is close to zero. For larger values
of $\mstopeff(\phi=0)$, as those associated with
$m_U^2 \geq 0$, there is a significant decrease of the order
parameter $v(T_c)/T_c$, with respect to the one used in
Eq.~(\ref{betaMSSM}), and hence the
relative finite temperature
QCD corrections may be larger than what is expected from
$\beta_{\rm MSSM}$ in Eq.~(\ref{betaMSSM}).
The results of Ref.~\cite{JoseR} confirm these expectations.

From the previous (qualitative) arguments one expects
that for $\mstopeff(\phi=0) \simeq 0$, from the validity of the
thermal perturbation theory, the upper bound on the Higgs mass
in the MSSM will be softened with
respect to that in the Standard Model by a factor $\sim
m_t/M_W$~\footnote{A similar observation was done in
Ref.~\cite{JRtalk}.}. In the Standard Model, lattice calculations
have shown that the electroweak phase transition is well
described by perturbation theory for Higgs masses $m_H\simlt
70$~GeV~\cite{nonpert}. Similarly we can expect
that in the MSSM, for the choice of supersymmetric
parameters rendering the phase transition much stronger than
in the Standard Model, the phase transition could be comfortably
well described up to Higgs masses $m_H\simlt M_Z$. Nevertheless,
it is clear that a rigorous proof of this statement would require
non-perturbative calculations, as previously stated.

In order to get $\mstopeff(\phi=0) \simeq 0$, the soft 
supersymmetry-breaking parameter $m_U^2$ must take negative values. Since
$T_c = {\cal{O}}(100\ {\rm GeV})$ and
$\Pi_R$ is of order $T^2$, Eq.~(\ref{polarization}),
relatively large negative values of $m_U^2$
must be phenomenologically acceptable. Such negative values of $m_U^2$ are
associated with the presence of charge- and
colour-breaking minima~\cite{CCB,CW}. As a conservative requirement,
it should demanded that the physical
vacuum state have lower energy than the color breaking minima.
We shall present
an analysis of the bounds obtained through such a requirement
in the next section.

\nsect{Colour-Breaking Minima at T = 0}

Let us first analyse the case of zero stop  mixing. In this case,
since $m_Q^2 \gg |m_U|^2$ the only fields that  acquire vacuum
expectation values are $\phi$ and
$U$. At zero temperature, the effective potential is given by
\be
V_{eff}(\phi,U) = -m_{\phi}^2 \phi^2 + \frac{\lambda}{2} \phi^4 + m_U^2 U^2
+ \frac{\widetilde{g}_3^2}{6} U^4 + \widetilde{h}_t^2 \sin^2\beta \phi^2 U^2
\label{effpphiu}
\ee
where $\lambda$ is the radiatively corrected
quartic coupling of the Higgs field,
with its corresponding dependence on the
top/stop spectrum through the one-loop
radiative corrections, $\widetilde{g}_3^2/3$ is the radiatively
corrected quartic self-coupling of the field $U$ and
$\widetilde{h}_t^2$ is the bi-bilinear $\phi-U$ coupling.
The latter couplings are well approximated by $\widetilde{g}_3 \simeq g_3$ and
$\widetilde{h}_t \simeq h_t$. For convenience, we shall define
\begin{equation}
\widetilde{m}^2_U = - m_U^2.
\end{equation}
The minimization of this potential leads to three extremes, at: {\bf (i)}
$\phi =0$, $U\neq0$; 
{\bf (ii)}~$U=0$, $\phi \neq 0$ and {\bf (iii)} $\phi \neq 0$, $U \neq 0$.
The corresponding expressions for the vacuum fields are:
\begin{equation}
\label{solutions}
\begin{array}{rllll}
{\bf (i)}&  U & = & 0, &
{\displaystyle \phi^2 = \frac{m_{\phi}^2}{\lambda}; }
\\ & & & & \\
{\bf (ii)}& \phi & = & 0, &
{\displaystyle U^2 = \frac{3 \widetilde{m}_U^2}{\widetilde{g}_3^2}; }
\\ & & & & \\
{\bf (iii)} & \phi^2 & = &
{\displaystyle \frac{m_{\phi}^2 - 3 \widetilde{m}_U^2 \widetilde{h}_t^2
\sin^2\beta/\widetilde{g}_3^2}{\lambda - 3 \widetilde{h}_t^4 \sin^4\beta/
\widetilde{g}_3^2}, }  &
{\displaystyle U^2 =
\frac{\widetilde{m}_{U}^2 - m_{\phi}^2  \widetilde{h}_t^2
\sin^2\beta/\lambda}{\widetilde{g}_3^2/3 -
\widetilde{h}_t^4 \sin^4\beta/\lambda}. }
\end{array}
\end{equation}
It is easy to show that the branch (iii) is continously connected with
branches (i) and (ii). It can also shown that the branch (iii) defines
a family of saddle-point solutions, the true (local) minima being defined by
(i) and (ii). Hence, the requirement of absence of a colour-breaking
minimum deeper than the physical one is given by
\begin{equation}
\widetilde{m}_U \leq
\left( \frac{m_H^2 \; v^2 \; \widetilde{g}_3^2}{12} \right)^{1/4}.
\label{boundmu}
\end{equation}
For a typical Higgs mass $m_H \simeq 70$ GeV, the bound on
$\widetilde{m}_U$ is of order 80 GeV.

In the case of stop mixing, $\widetilde{A}_t \neq 0$, the analysis is
more involved, since the three fields $Q$, $U$ and $\phi$ may
acquire vacuum expectation values. Due to the large hierarchy
between $m_Q^2$ and $m_U^2$, the vacuum expectation value of
$Q$ is always
small with respect to that of $U$, unless the mixing parameter
$\widetilde{A}_t$ is of order $m_Q$.  We shall hence define
\begin{equation}
\phi = \alpha \; U, \;\;\;\;\;\;\;\;\;\;\;\;\;\;\;\;\;
Q = \gamma \; U.
\end{equation}
The effective potential is given by
\begin{eqnarray}
V_{eff} & = & \left( -\widetilde{m}_U^2
- m_{\phi}^2 \alpha^2 + m_Q^2 \gamma^2 \right) U^2
+ 2 h_t \sin\beta \widetilde{A}_t \alpha \gamma U^3
\nonumber\\ \label{potab}
& + & U^4 \left[
\frac{\lambda}{2} \alpha^4 + h_t^2 \sin^2\beta
\alpha^2\left(1+
\gamma^2 \right)  + h_t^2 \gamma^2
+ \frac{g_3^2}{6} \left(1 - \gamma^2 \right)^2 \right. \\
& + & \left. \frac{1}{8} g^2 \gamma^2(\gamma^2+2\alpha^2\cos 2\beta)
+\frac{1}{72}g'^2(\gamma^2-4)(\gamma^2-4-6\alpha^2\cos 2\beta)
\right] \nonumber
\end{eqnarray}
As in Eq.~(\ref{effpphiu}),
we have not included 
in Eq.~(\ref{potab}) 
the radiative corrections
to the squark-Higgs and squark-squark couplings. Contrary to
what happens with the Higgs self-coupling, their tree-level
values are proportional to either the strong gauge coupling
or the top Yukawa coupling, and hence we expect the radiative corrections
to be suppressed by typical one-loop factors. Hence, for the purpose of
this paper, it is sufficient to keep their tree-level values. These
corrections must be included, however, if a more precise quantitative
study of the colour-breaking bounds is desired.

The effective potential can then be written as
\begin{equation}
V_{eff}(U,\gamma,\alpha) = F_1(\alpha,\gamma) \; U^2 + 
F_2(\alpha,\gamma) \; U^3
+ F_3(\alpha,\gamma) \; U^4
\end{equation}
where the expressions of the functions $F_i$ can be easily obtained from
Eq.~(\ref{potab}). 
In order to evaluate the depth of the color breaking minima,
we shall use the following procedure: We first minimize the potential with
respect to $U$. We find
\begin{equation}
U_{\rm min} =  \frac{-3 F_2 \; - \sqrt{9 F_2^2 - 32 F_1 F_3}}{8 F_3}\; ,
\end{equation}
where we have assumed that $\tilde{A}_t \geq 0$.
Inserting this solution into the effective potential, Eq.~(\ref{potab}),
we find 
\begin{equation}
\label{valpha}
V_{\rm min}(\gamma,\alpha) =
U_{\rm min}^2 \left( \frac{F_1}{3}  - U_{\rm min}^2 \frac {F_3}{3} \right)
\end{equation}
The resulting function of $\alpha$ and  $\gamma$,
Eq.~(\ref{valpha}), may be evaluated numerically. For each value of $\alpha$,
we have performed a scanning over $\gamma$, looking for the minimum value
of the effective potential, $V_{\rm min}(\alpha)$.
Fig.~1 shows the plot of $V_{\rm min}(\alpha)$ for
$m_Q=500$ GeV, $m_t=175$ GeV, $\tan\beta=1.7$ and
different values of $\widetilde{A}_t$. The value of the potential
at the minimum has been normalized to the
absolute value of the potential at the
physical expectation value $|V_{EW}|$,
so that $V_{\rm min}/|V_{EW}| = - 1$ for $\alpha \rightarrow \infty$.
Due to the effective potential structure, Eq.~(\ref{potab}), the
$\widetilde{A}_t$ effects are only relevant when the three fields
$U$, $Q$ and $\phi$ acquire a vacuum expectation value.
It is easy to show that larger values of $\widetilde{m}_U$
have the effect of inducing lower colour breaking minima for
both $\widetilde{A}_t = 0$ and $\widetilde{A}_t \neq 0$. Hence, in order
to obtain a conservative upper bound on $\widetilde{A}_t$,
we have chosen the (fixed) value of $\widetilde{m}_U$,
given by Eq.~(\ref{boundmu}), such
that the physical minimum (at $\alpha\rightarrow\infty$)
is degenerate with the colour breaking one at $\alpha = 0$.
We have explicitly checked that, as expected, 
for smaller values of $\widetilde{m}_U$,
the upper bound on $\widetilde{A}_t/m_Q$ moves to larger values.

For small and moderate values of $\widetilde{A}_t$
[$\widetilde{A}_t\simlt 430$ GeV in Fig.~1],
the saddle-point structure of the solutions (\ref{solutions})
with $U\neq0$ and $\phi \neq 0$ is clearly seen in the
figure as a maximum, while the only (degenerate) minima are
those at $\alpha=0,\; \infty$. Hence, so far
the condition (\ref{boundmu}) is fulfilled, the physical vacuum, with
$U = Q = 0$ is the true vacuum of the theory. This behaviour is preserved
for all values of $\widetilde{A}_t$ such that the present experimental
limit on the lightest stop is  fulfilled. Indeed, the upper bound on 
$\widetilde{A}_t$  is very close to the one obtained from the condition
of  avoiding a tachyon in the spectrum. In particular, for large values
of $\widetilde{A}_t$ [$\widetilde{A}_t\simgt 430$~GeV
in Fig.~1], a new global minimum with $\phi\neq Q\neq U\neq 0$
does appear, co-existing with the electroweak (local) minimum and the
saddle point (maximum). When a tachyonic state appears in the stop
spectrum [$\widetilde{A}_t \sim 450$ GeV] the electroweak minimum and
the saddle point collapse and turn into a single maximum.
As we shall show below, and as is 
clear from our discussion in the previous section,  Eq.~(\ref{totalE}),
large values of $\widetilde{A}_t$, close to the upper bound on this
quantity,  induce a large suppression of the
potential enhancement in the strength of the first-order phase 
transition through
the light top squark; they are hence  disfavoured from the point
of view of electroweak baryogenesis.

\nsect{Phase Transition Results}

As it follows from the discussion
in sections 2 and 3, larger values of
$\widetilde{m}_U$ can be helpful in inducing a strongly first-order
phase transition, but one must be careful about the presence of
charge- and
colour-breaking minima. Since the question of vacuum stability
is a delicate one, in this section we shall adopt the following
strategy: we shall in general
present results taking into account the vacuum stability constraint,
Eq.~(\ref{boundmu}). However,
the possibility that the physical vacuum is a metastable
state with a lifetime larger than the present age of the
Universe~\cite{tunnel}  can also be considered. In this
case, the bound Eq.~(\ref{boundmu}) would be inappropriate as a
phenomenological bound. In this article, we shall not address the question
of the vacuum state lifetime in quantitative terms. We shall limit ourselves
to also present the results obtained when the bound Eq.~(\ref{boundmu})
is ignored in the phenomenological analysis. However,
we shall always keep the constraint
\be
-\widetilde{m}_U^2+\Pi_R(T_c) > 0.
\label{stability}
\ee
Indeed, if Eq.~(\ref{stability}) were not fulfilled, the Universe
would be driven to a charge- and colour-breaking minimum
at $T\geq T_c$. Moreover, since the
transition to the color breaking minimum is first order, one
should also require the critical temperature for the transition to
this minimum, $T_c^U$, to be below $T_c$. Because of 
the strength of
the stop coupling to the gluon and squark fields, one should
expect this transition to be more strongly first order than the
electroweak  one.  

We shall assume that, as happens at zero temperature,
it is sufficient to analyse the behaviour of the potential 
in the direction  $U \neq 0$, $\phi = Q = 0$, 
to determine the conditions that assure
the stability of the physical vacuum at finite temperature. 
In order to get a quantitative bound on the
mass parameter $\widetilde{m}_U$,
the effective finite temperature potential for
the $U$ field must be analysed. 
For this purpose,
it is useful to compute  the particle
spectrum in a non-vanishing $U$-field background. The most
relevant masses are: \\
a) The hypercharge ($B$) gauge boson with squared mass
$8 g'^2 U^2/9$. \\
b) Four gluons with squared masses $g_3^2 U^2/2$
and one gluon with squared mass  $2 g_3^2 U^2/3$. \\
c) Five squarks (would-be Goldstones) with
squared masses $-\widetilde{m}_U^2
+ (g_3^2+4 g'^2/3) U^2/3$ and one with squared mass
$-\widetilde{m}_U^2 + (g_3^2+4 g'^2/3) U^2$.\\
d) Four scalar (Higgs-left-handed squark) states with squared masses \\
$-m_H^2/2 + \left[h_t^2 \sin^2\beta
\left( 1 - \widetilde{A}_t^2/m_Q^2 \right)
-|\cos 2\beta|g'^2/3 \right] U^2$.\\
e) Two Dirac fermion states (left quark-Higgsino) with squared masses
$\mu^2 + h_t^2 U^2$ and two Majorana fermion states
(right top-bino) with masses
$\sqrt{8 g'^2/9 U^2 + (M_1/2)^2} \pm M_1/2$. \\
Hence, the finite-temperature
effective potential for the $U$ field is given by
\begin{equation}
V_U = \left(-\widetilde{m}_U^2 + \gamma_U T^2 \right) U^2 -
T E_U U^3 + \frac{\lambda_U}{2} U^4,
\label{upot}
\end{equation}
where
\begin{eqnarray}
\gamma_U & \equiv & \frac{\Pi_R(T)}{T^2} \simeq
\frac{4 g_3^2}{9} + \frac{h_t^2}{6}\left[ 1 +
\sin^2\beta (1 - \widetilde{A}_t^2/m_Q^2)
\right] ; \;\;\;\;\;\;\;\;\;\;\;\;\;\;
\lambda_U \simeq \frac{g_3^2}{3}
\nonumber\\
\label{totalEu}
E_U & \simeq &  \left[\frac{\sqrt{2} g_3^2}{6 \pi} \left( 1 +
\frac{2}{3\sqrt{3}} \right) \right] \\
& + &
\left\{ \frac{g_3^3}{12\pi}\left(\frac{5}{3\sqrt{3}} + 1\right)
+ \frac{h_t^3 \sin^3\beta (1-\widetilde{A}_t^2/m_Q^2)^{3/2}}{3 \pi} \right\}.
\nonumber
\end{eqnarray}
The
contribution to $E_U$ inside the squared brackets comes
from the transverse gluons, $E_U^g$, while the one inside the curly
bracket comes from the squark and Higgs contributions
[for simplicity of  presentation,
we have not written explicitly the small hypercharge contributions
to $E_U$ and $\gamma_U$.].  In the above, we have ignored the gluino and
left-handed squark contributions since they are assumed to be heavy and,
as we explained above, their contributions to the finite
temperature effective potential is Boltzmann-suppressed.

The difference between $T_0^U$, the temperature at which
$\mstopeff(\phi = 0)  = 0$, and $T_c^U$, is given by
\begin{equation}
T_c^U = \frac{T_0^U}{\sqrt{1 - E_U^2/(2\lambda_U\gamma_U)}}.
\end{equation}
In order to assure a transition from the $SU(2)_L \times U(1)_Y$
symmetric minimum to the physical one at $T = T_c$, we
should replace the condition (\ref{stability}) by the condition
\begin{equation}
-\widetilde{m}_U^2 + \Pi_R(T) > \widetilde{m}_U^2
\frac{\epsilon}{1-\epsilon} \simeq \widetilde{m}_U^2 \epsilon,
\label{stability2}
\end{equation}
with $\epsilon = E_U^2/2\lambda_U\gamma_U$, a small number.
In Eq. (\ref{totalEu})
we have written the value of $E_U$ that would be obtained if
the field-independent effective thermal mass terms of the
squark and Higgs fields were exactly
vanishing at the temperature $T$. Although for values of
$\widetilde{m}_U^2$, which induce a large cubic term in the
Higgs potential, $T_c$ is actually
close to the temperature at which these masses vanish, an
effective screening is always present. 
In the following, we shall require the stability condition,
Eq.~(\ref{stability2}), while using the value of $E_U$
given in Eq.~(\ref{totalEu}). We shall also show the result that
would be obtained if only the gluon contributions to $E_U$,
$E_U^{g}$, would
be considered. The difference between the two results 
quantifies the uncertainty in $E_U$  due to the fact that
the effective thermal masses of the squark and the Higgs
fields are actually partially screened at $T_c$.

Let us first present the results for zero mixing.
Fig.~2 shows the order parameter $v(T_c)/T_c$
for the phase transition as a function of the running light stop mass,
for $\tan\beta = 2$, $m_Q = 500$ GeV
and $m_t = 175$ GeV. For these parameters, the Higgs mass
$m_H \simeq 70$ GeV, a result that depends weakly on $\widetilde{m}_U$.
We see that for smaller (larger) values of
$\mstop$ ($\widetilde{m}_U$),  $v(T_c)/T_c$
increases in accordance with the discussion of section 2.
We have marked with a diamond the lower bound on the stop mass coming from
the bound on colour-breaking vacua at $T=0$, Eq.~(\ref{boundmu}).
The cross and the star denote the bounds that would be obtained
by requiring  condition (\ref{stability2}), while using
the total and gluon-induced trilinear coefficients,
$E_U$ and $E_U^g$, respectively.
We see that the light stop effect is
maximum for values of $\widetilde{m}_U^2$ such that
condition (\ref{stability2}) is saturated,
which leads to values of  $\mstop \simeq 140$ GeV
($\widetilde{m}_U\simeq 90$ GeV) and $v(T_c)/T_c \simeq 1.75$.
To preserve
Eq. (\ref{boundmu}) demands slightly larger stop mass values. Still,
there is a large region of parameter space for which
$v(T_c)/T_c \geq 1$ and is not in conflict with any phenomenological
constraint.

Figure 3 shows the results for zero mixing and
$m_Q = 500 $ GeV as a function of $\tan\beta$
and for the values of $\widetilde{m}_U$ such that the maximum
effect on $v(T_c)/T_c$ is achieved.
We also plot in this figure the corresponding values of the stop and Higgs
masses. As in Fig.~2,
the solid [dashed] line
represents the result when the bound (\ref{boundmu})
[the stability bound of Eq.~(\ref{stability2})] is preserved.
We see that $v(T_c)/T_c$
increases for lower values of $\tan\beta$, a change mainly associated with
the decreasing value of the Higgs mass or, equivalently, of the Higgs
self-coupling. For values of $\tan\beta \simeq 2.7$, $v(T_c)/T_c \simeq 1$,
and hence the value of the Higgs mass yields the upper bound
consistent with electroweak baryogenesis.
This bound is approximately given by $m_H \simeq 80$ GeV. If the bound
on color breaking minima, Eq.~(\ref{boundmu}), is ignored, the upper
bound on $m_H$ is close to 100 GeV, in accordance with our qualitative
discussion  of section 1.

Due to the logarithmic dependence of $m_H$ on $m_Q$, larger values of
$m_Q$ have the effect of enhancing the Higgs mass values. It turns out
that, for zero mixing, the results for $v(T_c)/T_c$ depend on the Higgs mass
and on the value of $m_U$, but not on the specific value of $m_Q$. Hence,
different values of $m_Q$ have the only effect of shifting (up or down) the
preferred values of $\tan\beta$. In particular,
the fixed-point solution, which
corresponds to values of $\tan\beta \simeq 1.6$ for $m_t \simeq 175$ GeV,
leads to values of $m_H \geq 65$ GeV and
$v(T_c)/T_c \simgt 1$ so far $m_Q$ is above 750 GeV and below a
few TeV.

Finally, let us discuss the effect of mixing in the stop sector.
For fixed  values of $m_Q$
and $\tan\beta$, increasing the values of
$\widetilde{A}_t$ has a negative effect
on the strength of the first-order phase transition for three reasons. First,
large values of $\widetilde{A}_t$ lead to larger values of the Higgs mass
$m_H$. Secondly, as shown in Eq.~(\ref{totalE}) they suppress the stop
enhancement of the cubic term. Finally, there is an indirect effect associated
with the constraints on the allowed values for
$\widetilde{m}_U$. This has to do
with the fact that for larger values of
$\widetilde{A}_t$, the phase transition
temperature increases, rendering
more difficult an effective suppression of
the effective mass $\mstopeff$, Eq.~(\ref{effmass}).
Of course, this third reason is absent if the bound (\ref{boundmu})
is ignored.
As we have shown above, for zero mixing the bounds (\ref{orderp}),
(\ref{boundmu})  and  (\ref{stability2})
are only fulfilled for values
of the stop mass larger than approximately 140 GeV.
Light stops, with masses $\mstop \simlt 100$ GeV,
can only be consistent with these constraints for
larger values of the mixing mass parameter $\widetilde{A}_t$.
This can be relevant
for physical processes, which demand the presence of
such light sparticles in the spectrum.
For instance, it is important in getting corrections
to $R_b$~\cite{precision,CW,lowtb}.

Figure 4 shows the result for $v(T_c)/T_c$ as a function of $\widetilde{A}_t$
for $\tan\beta = 1.7$,
$m_Q = 500$~GeV, and values of $m_U$ such that the maximal light
stop effect is achieved.
The same conventions as in Fig.~3 have been used.
Due to the constraints on $\widetilde{m}_U$,
light stops with $\mstop \simlt M_W$ may only be obtained for values of
$\widetilde{A}_t \simgt 0.6\; m_Q$. For these values of $\widetilde{A}_t$,
however, the phase-transition temperature is large and
induces large values of $\mstopeff$,
for all values of $\widetilde{m}_U$
allowed by Eq.~(\ref{boundmu}).
In Fig.~4, we have chosen the parameters such that they lead to the
maximum value of the mixing parameter $\widetilde{A}_t/m_Q$ consistent
with $v(T_c)/T_c \geq 1$ and the Higgs mass bound. As we mentioned
above, lower values of $\tan\beta$ can be achieved for larger values
of $m_Q$. Since the stop spectrum depends only slightly on $\tan\beta$,
we obtain that, as
far as the bounds on color breaking minima are preserved, the mixing
effects are not very helpful to obtain lower stop masses
compatible with a sufficiently strong first order phase transition.
If the weaker bound, Eq.~(\ref{stability2}), were required
(thin and thick dashed lines in Fig.~4), light stops,
with masses of order $M_Z$  would not be in conflict with
electroweak baryogenesis.

\nsect{Conclusions}

In this article we  show  that, contrary to what was suggested
by previous analyses,
there are large regions of phenomenologically acceptable parameter
space, that are consistent with the present experimental bound
on the Higgs mass and with a sufficiently strong 
electroweak first-order phase
transition, Eq.~(\ref{orderp}).
This region of parameter space is associated
with low values of $\tan\beta$, low values of the lightest stop
mass, $\mstop \simlt m_t$,
and low values of the Higgs mass, $m_H \simlt M_Z$.
It can hence be tested by experimental Higgs and stop
searches at the Tevatron  and LEP2  colliders. Interestingly enough,
this region is also consistent with the unification of the bottom
and $\tau$ Yukawa couplings at the grand unification scale, and
consequently with the quasi-infrared fixed point solution
for the top quark mass. The hierarchy of soft supersymmetry breaking
parameters $m_Q^2 \gg m_U^2$ is naturally
obtained at the fixed-point if supersymmetry is broken in a
hidden sector. Furthermore, as has been discussed in Ref.~\cite{CW},
negative values of $m_U^2$ are associated with non-universal
boundary conditions for the scalar soft supersymmetry-breaking
terms at the scale $M_{GUT}$. For these values of $m_U^2$,
the bounds on
the colour breaking minima are decisive in defining the allowed
parameter space and, for a top quark mass $m_t \simeq 175$ GeV,
light stop masses below  130 GeV turn to be
disfavoured. If these constraints are ignored, while assuming that
we live in a metastable vacuum,
light stops, with masses of the order of the $Z^0$ mass become
consistent with a strongly first-order electroweak phase transition.

Three additional remarks are in order: i) First, we have always considered
the case of very large $m_Q$. A stronger first-order phase transition
may be obtained by considering values of $m_Q$ such that the left
handed stop finite temperature contribution is not negligible.
Since the Higgs mass value may be controlled through $\tan\beta$,
the strongest bounds on $m_Q$ come from preserving a good fit
to the electroweak precision measurements. As we discussed above,
for the experimentally preferred values of the top quark mass,
large values of $m_Q$ are preferred. ii) We have only analysed the
case of large values of the CP-odd Higgs mass. For the CP-violating
effects associated with the supersymmetric particles to lead to
an efficient baryon generation at the electroweak phase transition,
the ratio of vacuum expectation values must change along the
bubble walls \cite{CPMSSM}.
This in turn means that the CP-odd mass cannot be much
larger than the critical temperature. It is difficult to derive a
quantitative upper bound on the CP-odd Higgs mass from these considerations.
However, since there is no significant change of the order parameter
$v(T_c)/T_c$ up to CP-odd Higgs masses as low as $\simeq 2 T_c$,
we do not expect a relevant change of the allowed parameter space
with respect to the one found in the present analysis. iii)
Throughout this paper we have ignored higher-loop corrections.
% In
% section 2, we have argued that the most relevant corrections are
% expected to be the QCD ones.
These corrections tend to make the phase transition
more strongly first order \cite{JoseR} and enlarge
the allowed  parameter space.
A non-perturbative study will be useful to check the
validity of the perturbative bounds on the MSSM parameters
consistent with  electroweak baryogenesis.

\begin{center}
{\bf Acknowledgements}
\end{center} \noindent
We would like to thank A. Brignole,
Z. Fodor, M. Pietroni,
N. Rius, M. Shaposhnikov,
C. Wetterich, F. Zwirner and, in particular,
J.R. Espinosa for many  useful discussions.

\newpage
%%%%%%%%%%%%%%%%%%%%%%%%figure%%%%%%%%%%%%%%%%%%%%%%%%
\begin{figure}
%\psdraft
%\centerline{
\psfig{figure=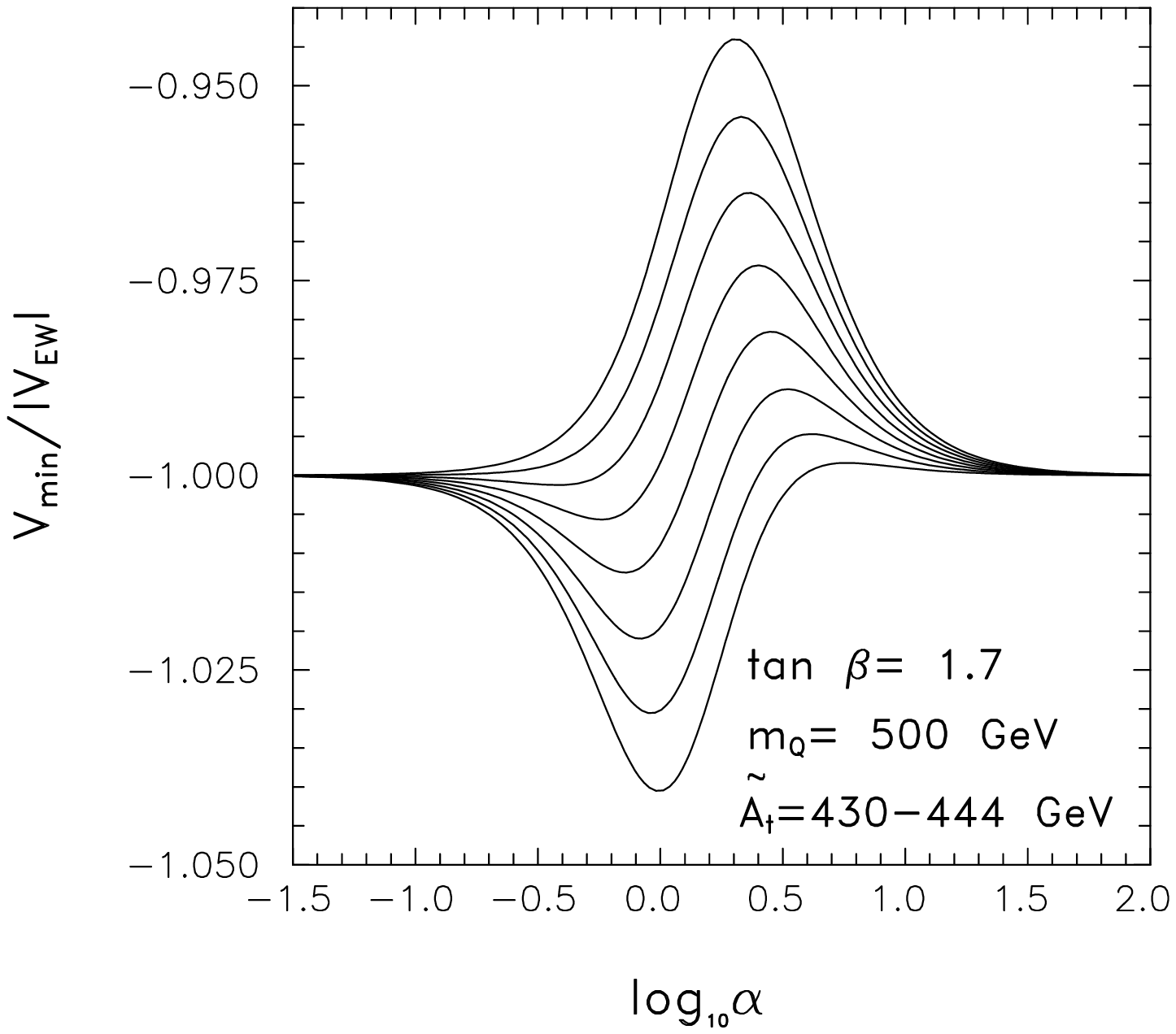,width=19.0cm,height=19.0cm}
\vspace{1.0truecm} 
% bbllx=4.5cm,bblly=.cm,bburx=14.cm,bbury=13cm}}
\caption{Plot of $V_{\rm min}/|V_{EW}|$ for $m_t=175$ GeV, $m_Q=500$ GeV,
$\tan\beta=1.7$ and $\widetilde{A}_t$=430 GeV 
[upper curve]--444 GeV [lower curve], step=2 GeV.}
\end{figure}
%%%%%%%%%%%%%%%%%%%%%%%%figure%%%%%%%%%%%%%%%%%%%%%%%%

%%%%%%%%%%%%%%%%%%%%%%%%figure%%%%%%%%%%%%%%%%%%%%%%%%
\begin{figure}
%\psdraft
%\centerline{
%\psfig{figure=fig2.ps,
\psfig{figure=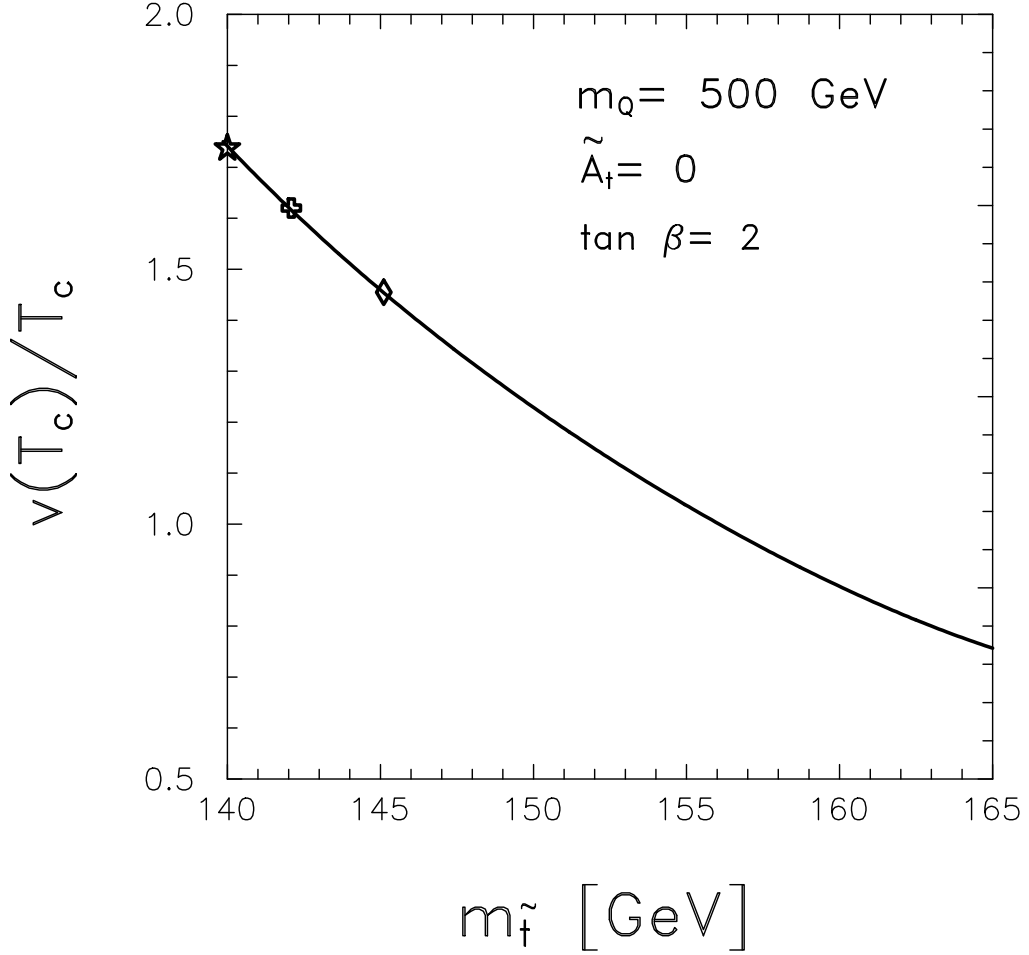,width=19.0cm,height=19.0cm}
\vspace{1.0truecm} 
%height=13cm,bbllx=4.5cm,bblly=.cm,bburx=14.cm,bbury=13cm}}
\caption{Plot of $v(T_c)/T_c$ as a function of $\mstop$ for $m_Q$ and
$m_t$ as in Fig.~1, $\widetilde{A}_t=0$ and $\tan\beta=2$. The
diamond [cross, star] denotes the value of $\widetilde{m}_U$
for which the bound, Eq.~(3.4) [Eq.~(4.5) with
$E_U$ given by Eq.~(4.3), Eq.~(4.5) with
$E_U = E_U^g$] is saturated.}
\end{figure}
%%%%%%%%%%%%%%%%%%%%%%%%figure%%%%%%%%%%%%%%%%%%%%%%%%

%%%%%%%%%%%%%%%%%%%%%%%%figure%%%%%%%%%%%%%%%%%%%%%%%%
\begin{figure}
%\psdraft
%\centerline{
%\psfig{figure=fig3.ps,
\psfig{figure=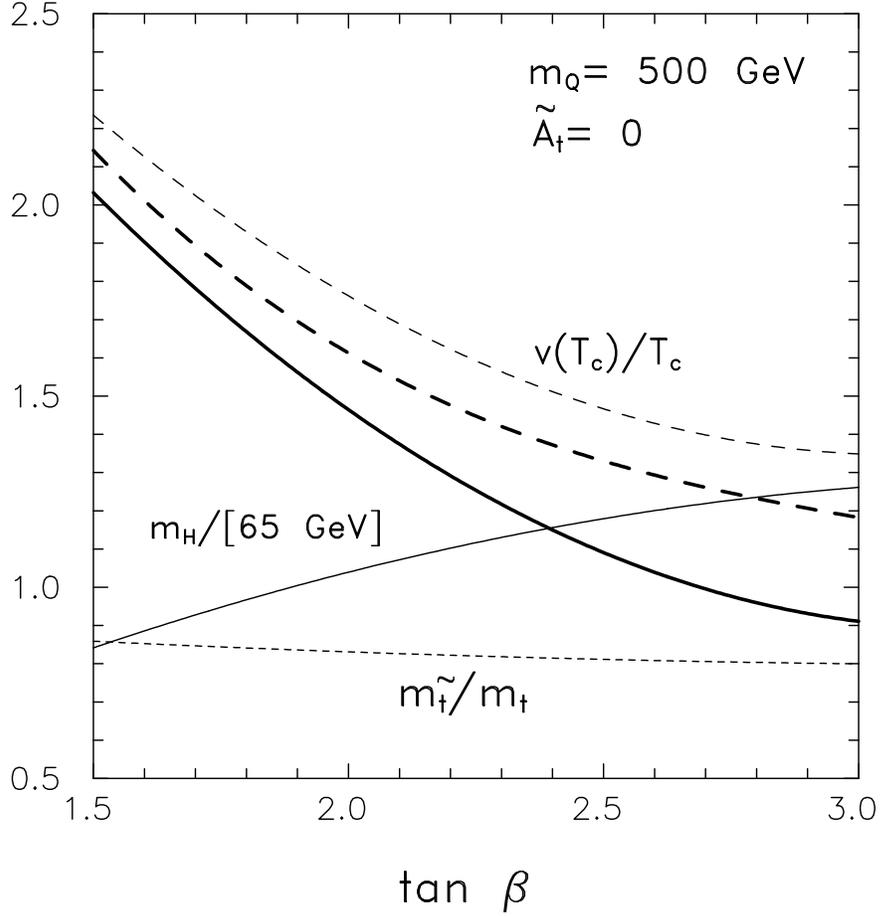,width=19.0cm,height=19.0cm}
\vspace{1.0truecm} 
%height=13cm,bbllx=4.5cm,bblly=.cm,bburx=14.cm,bbury=13cm}}
\caption{Plot of $v(T_c)/T_c$ as functions of $\tan\beta$
for $m_Q$ and $\widetilde{A}_t$ as in Fig.~2, and $m_U$ saturating
Eq.~(3.4) [solid] and Eq.~(4.5) [thick dashed line
for $E_U$ given by Eq.~(4.3) and thin dashed line for
$E_U = E_U^g$].
The additional thin
lines are plots of $m_H$ in units of 65 GeV [solid] and $\mstop$
in units of $m_t$ [short-dashed],
corresponding to the values of $\widetilde{m}_U$ associated
with the solid line.}
\end{figure}
%%%%%%%%%%%%%%%%%%%%%%%%figure%%%%%%%%%%%%%%%%%%%%%%%%

%%%%%%%%%%%%%%%%%%%%%%%%figure%%%%%%%%%%%%%%%%%%%%%%%%
\begin{figure}
%\psdraft
%\centerline{
%\psfig{figure=fig4.ps,
\psfig{figure=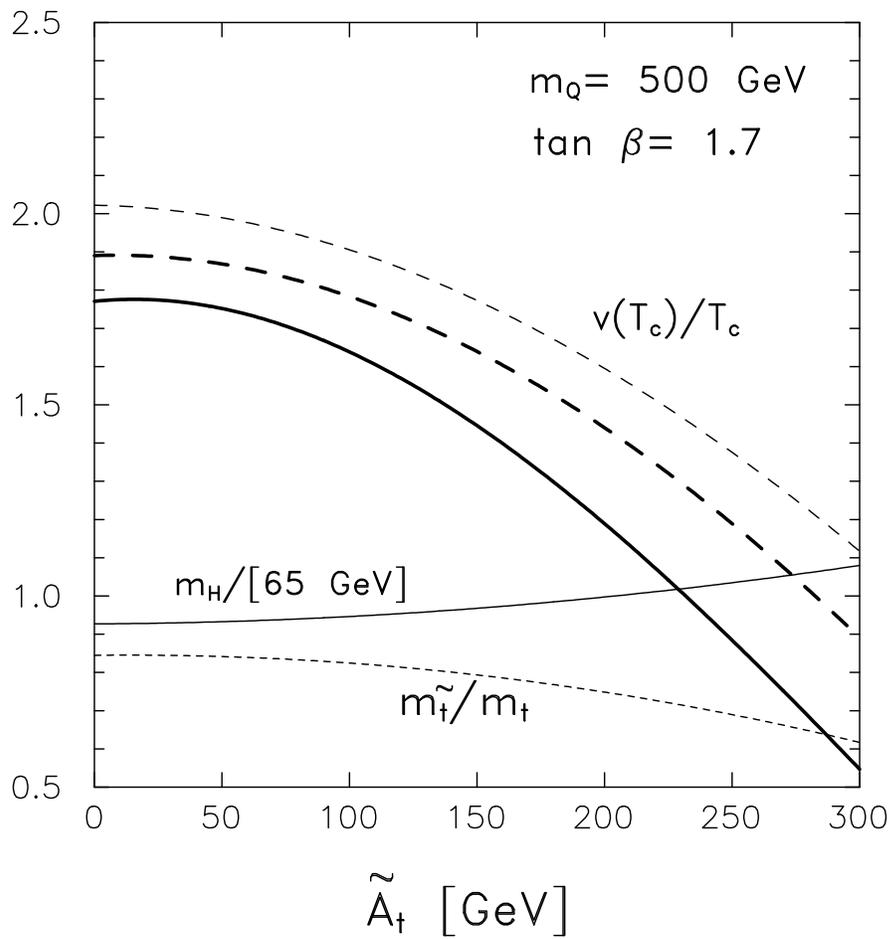,width=19.0cm,height=19.0cm}
\vspace{1.0truecm} 
%height=13cm,bbllx=4.5cm,bblly=.cm,bburx=14.cm,bbury=13cm}}
\caption{The same as in Fig.~3, but as functions of $\widetilde{A}_t$, for
$\tan\beta=1.7$.}
\end{figure}
%%%%%%%%%%%%%%%%%%%%%%%%figure%%%%%%%%%%%%%%%%%%%%%%%%

\newpage
\begin{thebibliography}{99}
%
\bibitem{baryogenesis} A.D.~Sakharov, \JETPL{91B}{67}{24}
%
\bibitem{anomaly}G.~t'Hooft, \PRL{37}{76}{8}; \PRD{14}{76}{3432}
%
\bibitem{sphalerons} P.~Arnold and L.~McLerran, \PRD{36}{87}{581};
and {\bf D37} (1988) 1020; S.Yu~Khlebnikov and M.E.~Shaposhnikov,
\NPB{308}{88}{885};
F.R. Klinkhamer and N.S. Manton, \PRD{30}{84}{2212};
B. Kastening, R.D. Peccei and X. Zhang, \PLB{266}{91}{413};
L.~Carson, Xu~Li, L.~McLerran and R.-T.~Wang, \PRD{42}{90}{2127};
M.~Dine, P.~Huet and R.~Singleton Jr., \NPB{375}{92}{625}
%
\bibitem{reviews} For recent reviews, see:
A.G. Cohen, D.B. Kaplan and A.E. Nelson,
{\it Annu. Rev. Nucl. Part. Sci.} {\bf 43} (1993) 27;
M. Quir\'os, \HPA{67}{94}{451}; V.A.~Rubakov and M.E.~Shaposhnikov,
preprint CERN-TH/96-13 [hep-ph/9603208]
%
\bibitem{first} M. Shaposhnikov, 
{\it JETP Lett.} 44 (1986) 465; \NPB{287}{87}{757} and
{\bf B299} (1988) 797
%
\bibitem{improvement} M.E. Carrington, \PRD{45}{92}{2933};
M. Dine, R.G. Leigh, P. Huet, A. Linde and D. Linde, \PLB{283}{92}{319};
\PRD{46}{92}{550}; P. Arnold, \PRD{46}{92}{2628};
J.R. Espinosa, M. Quir\'os and F. Zwirner, \PLB{314}{93}{206};
W. Buchm\"uller, Z. Fodor, T. Helbig and D. Walliser, \AP{234}{94}{260}
%
\bibitem{twoloop} J.~Bagnasco and M.~Dine, \PLB{303}{93}{308};
P. Arnold and O. Espinosa, \PRD{47}{93}{3546}; Z. Fodor and A. Hebecker,
\NPB{432}{94}{127}
%
\bibitem{nonpert} K. Kajantie, K.~Rummukainen
and M.E.~Shaposhnikov, \NPB{407}{93}{356};
 Z. Fodor, J. Hein, K. Jansen, A. Jaster and
I. Montvay, \NPB{439}{95}{147};
K.~Kajantie, M.~Laine, K.~Rummukainen and
M.E.~Shaposhnikov, preprint CERN-TH/95-263 [hep-lat/9510020];
K.~Jansen, preprint DESY 95-169 (September 1995).\\
For an alternative approach, see: B. Bergerhoff and
C. Wetterich, \NPB{440}{95}{171} and references therein
%
\bibitem{CPSM} G.R.~Farrar and M.E.~Shaposhnikov, \PRL{70}{93}{2833} and
({\bf E}): {\bf 71} (1993) 210; M.B.~Gavela. P.~Hern\'andez, J.~Orloff,
O.~P\`ene and C.~Quimbay, \MPLA{9}{94}{795};
\NPB{430}{94}{382}; P.~Huet and E.~Sather, \PRD{51}{95}{379}
%
\bibitem{CPMSSM} M.~Dine, P.~Huet, R.~Singleton Jr. and L.~Susskind,
\PLB{257}{91}{351}; A.~Cohen and A.E.~Nelson, \PLB{297}{92}{111};
P.~Huet and A.E.~Nelson, preprint UW/PT 95-07 [hep-ph/9506477]
%
\bibitem{SPCP} D.~Comelli and M.~Pietroni, \PLB{306}{93}{67};
J.R.~Espinosa, J.M.~Moreno and M.~Quir\'os, \PLB{319}{93}{505};
D.~Comelli, M.~Pietroni and A.~Riotto, \NPB{412}{94}{441}
%
\bibitem{early} G.F. Giudice, \PRD{45}{92}{3177};
S. Myint, \PLB{287}{92}{325}
%
\bibitem{mariano1} J.R. Espinosa, M. Quir\'os and F. Zwirner,
\PLB{307}{93}{106}
%
\bibitem{mariano2} A. Brignole, J.R. Espinosa, M. Quir\'os and F. Zwirner,
\PLB{324}{94}{181}
%
\bibitem{precision} G. Altarelli, R. Barbieri and F. Caravaglios,
\PLB{314}{93}{357}; \NPB{405}{93}{3};
J.D.~Wells, C.~Kolda and G.L.~Kane,
\PRL{338}{94}{219}; D.~Garc\'{\i}a, R.~Jim\'enez and J. Sol\`a,
\PLB{347}{95}{309} and (E) {\bf B351} (1995) 602;
P.~Chankowski and S.~Pokorski, in
{\it Beyond the Standard Model IV}, Granlibakken, Tahoe City, CA,
13--18 December 1994, eds. J.F.~Gunion, T.~Han and J.~Ohnermus, 
(World Scientific, Singapore, 1995) pp. 233-242; A.~Dabelstein,
W.~Hollik and W.~M\"osle, in {\it Perspectives for Electroweak
Interactions in $e^+e^-$ Collisions}, Ringberg Castle, Germany,
5-8 February 1995, B. Kniehl ed. (World Scientific,
Singapore, 1995) pp. 345-361
%
\bibitem{COPW}
M. Carena, M. Olechowski, S. Pokorski and C.E.M. Wagner,
\NPB{419}{94}{213};
V. Barger, M.S. Berger and P. Ohmann, \PRD{49}{94}{4908};
B. Ananthanarayan, K.S. Babu and Q. Shafi, \NPB{428}{94}{19};
G. Kane, C. Kolda, L. Roszkowski and J.D. Wells, \PRD{50}{94}{3498};
J. Gunion and H. Pois, \PLB{329}{94}{736}; G.K. Leontaris and
N.D. Tracas, \PLB{336}{94}{194}; W. de Boer, R. Ehret and D.I.
Kazakov, \PLB{329}{94}{736}
%
\bibitem{CW} M. Carena and C.E.M. Wagner, \NPB{452}{95}{45}
%
\bibitem{Higgs}
Y. Okada, M. Yamaguchi and T. Yanagida, {\it Prog. Theor. Phys.}
{\bf 85} (1991) 1; H.E. Haber and R. Hempfling, \PRL{66}{91}{1815}
%
\bibitem{BERZ} J. Ellis, G. Ridolfi and F. Zwirner and 
{\bf B257} (1991) 83;
\PLB{262}{91}{477}; A. Brignole, J. Ellis, G. Ridolfi and F. Zwirner,
\PLB{271}{92}{123}; A. Brignole, \PLB{281}{92}{284}
%
\bibitem{CEQW} J.A. Casas, J.R. Espinosa, M. Quir\'os
and A. Riotto, \NPB{436}{95}{3};
M.~Carena, J.R.~Espinosa, M.~Quir\'os and C.E.M.~Wagner,
\PLB{335}{95}{209}; M.~Carena, M.~Quir\'os and C.E.M.~Wagner,
CERN preprint CERN-TH/95-157 [hep-ph/9508343], to appear in {\it Nucl.
Phys.} {\bf B}; H. Haber, R. Hempfling and H. Hoang, CERN preprint
CERN-TH/95-216 (to appear);
M.~Carena, P.~Zerwas and the Higgs Physics Working
Group, in
Vol.~1 of {\it Physics
at LEP2}, G.~Altarelli, T.~Sj\"ostrand and
F.~Zwirner, eds., Report CERN 96-01, Geneva (1996), to appear
%
\bibitem{JoseR} J.R. Espinosa, DESY and IEM-FT-122/95 preprint (to appear)
%
\bibitem{JRtalk} J.R.~Espinosa, in {\it Electroweak Physics and the
Early Universe}, eds. J.C.~Romao and F.~Freire, Plenum Press,
New York, 1994, p. 93
%
\bibitem{CCB} J.M.~Fr\`ere, D.R.T.~Jones and S.~Raby, \NPB{222}{83}{11};
L.~Alvarez-Gaum\'e, J.~Polchinski and M.B.~Wise, \NPB{221}{83}{495};
C.~Kounnas, A.B.~Lahanas, D.V.~Nanopoulos and M.~Quir\'os,
\NPB{236}{84}{438}; J.P.~Derendinger and C.A.~Savoy, \NPB{237}{84}{307};
J.F.~Gunion, H.E.~Haber and M.~Sher, \NPB{306}{88}{1};
H.~Komatsu, \PLB{215}{88}{323}; P.~Langacker and N.~Polonsky,
\PRD{50}{94}{2199}; A.J.~Bordner, Kyoto preprint KUNS-1351, HE(TH) 95/11
[hep-ph/9506409]; J.A.~Casas, A.~Lleyda and C.~Mu\~noz, Madrid preprint
FTUAM 95/11, IEM-FT-100/95 [hep-ph/9507294]
%
\bibitem{tunnel} M.~Claudson, L.J.~Hall and I.~Hinchliffe, \NPB{228}{83}
{501}; A.~Riotto and E.~Roulet, preprint FERMILAB-Pub-95/400-A,
SISSA-163/95/EP [hep-ph/9512401]; A.~Kusenko, P.~Langacker and
G.~Segr\'e, preprint UPR-677-T [hep-ph/9602414]
%
\bibitem{lowtb} J.D. Wells and G.L. Kane, preprint SLAC-PUB-7038
[hep-ph/9510372]; J. Feng, N. Polonsky and S. Thomas,
preprint SLAC-PUB-95-7050 [hep-ph/9511324]; J. Ellis, J.L. Lopez
and D.V. Nanopoulos, preprint CERN-TH/95-314 [hep-ph/9512288];
A. Brignole, F. Feruglio and F. Zwirner, preprint CERN-TH/95-340
[hep-ph/9601293]; E.H. Simmons and Y. Su, preprint BU-HEP-4,
[hep-ph/9602267]; P.H. Chankowski and S. Pokorski, preprint
IFT-96/6 [hep-ph/9603310]
\end{thebibliography}
\end{document}